\documentclass[11pt,twoside]{article}
\usepackage{asp2004}
\usepackage{psfig}
\usepackage{epsf}
\usepackage{graphics}
\usepackage{lscape}
\usepackage{natbib}
\def\edcomment#1{\iffalse\marginpar{\raggedright\sl#1\/}\else\relax\fi}
\reversemarginpar

\begin{document}
\title{The Supernova Legacy Survey}
\author{Mark Sullivan (for the SNLS collaboration$^1$)}
\affil{Dept. of Astronomy and Astrophysics, University of Toronto, 60 St. George St., Toronto M5S3H8, Canada}

\begin{abstract}
  Type Ia supernovae (SNe\,Ia) currently provide the most direct
  evidence for an accelerating Universe and for the existence of an
  unknown ``dark energy'' driving this expansion. The 5-year Supernova
  Legacy Survey\footnote{see \texttt{http://cfht.hawaii.edu/SNLS/}} (SNLS) 
  will deliver $\sim$1000 SN\,Ia detections with
  well-sampled $g'r'i'z'$ light curves.  Using this definitive
  dataset, we will obtain a precise measurement of the cosmological
  parameters ($\Omega_{m}$, $\Omega_{\lambda}$); our goal is to
  determine the cosmological equation of state parameter ``$w$'' to a
  statistical precision of $\pm0.07$, testing theories for the origin
  of the universal acceleration. In this proceeding, we summarise the
  progress made during the first full year of the survey operation.
\end{abstract}
\thispagestyle{plain}

\section{Introduction}

The Hubble diagram for high-redshift Type Ia supernovae (SNe\,Ia)
provides the most direct current measurement of the expansion history
of the universe -- and hence the most direct evidence for an
accelerating expansion.  The ``first generation'' of SN\,Ia cosmology
work developed a systematic approach to this measurement (Riess et~al.
1998; Perlmutter et~al. 1999) that led to astonishing results ruling
out a flat, matter-dominated Universe. This indicated the presence of
a new, unaccounted-for ``dark energy'' driving the cosmic acceleration.

One of the most pressing questions in cosmology now is: ``What is the
dark energy that causes this acceleration?''. There are fundamental
differences between a Cosmological Constant and other proposed forms
of dark energy (see Peebles \& Ratra 2003 for a comprehensive review).
The distinction can be addressed by measuring the dark energy's
average equation-of-state, $<$$w$$>$$=$$<$$p/\rho$$>$, where
$w$$=$$-1$ corresponds to a Cosmological Constant -- current
measurements of this parameter (e.g., Knop et~al.  2003; Tonry et~al.
2003; Riess et~al.  2004) are consistent with a very wide range of
dark energy theories.  The importance of improving measurements to the
point where $w$$=$$-1$ could be excluded has led to a
second-generation of SN cosmology studies: large multi-year,
multi-observatory programs benefiting from major commitments of
dedicated time.  These ``rolling searches'' find and follow SNe over
many consecutive months of repeated wide-field imaging, with redshifts
and SN type classification from coordinated spectroscopy.  The
Canada-France-Hawaii Telescope (CFHT) Legacy Survey is one such
ambitious repeat-imaging wide-field survey conducted in 4 SDSS filters
($g'r'i'z'$), utilising an imager field three times larger than used
in the next largest survey, with around twice as much time devoted to
the survey. The five-year CFHT ``Supernova Legacy Survey'' (SNLS) will
provide the biggest improvement in the determination of the dark
energy parameters achievable over the next decade, using an
order-of-magnitude larger statistical sample (i.e.  $\sim$700) of SNe
in the redshift range $z$$=$0.3-0.9 where $w$ is best measured.  With
this sample, we aim to answer the key question: Is the dark energy
something other than Einstein's $\Lambda$?

\section{Survey overview}

\begin{table}
\begin{center}

\caption{SNLS field locations\label{fieldpositions}}
\smallskip
{\small
\begin{tabular}{cllc}
\tableline
\noalign{\smallskip}
Field & RA\,(J2000) & DEC\,(J2000)&Other data\\
\noalign{\smallskip}
\tableline
\noalign{\smallskip}
D1 & 02:26:00.00 & $-$04:30:00.0 & XMM-Deep, VIMOS, SWIRE, GALEX\\
D2 & 10:00:28.60 & +02:12:21.0 & COSMOS/ACS, VIMOS, SIRTF, GALEX\\
D3 & 14:19:28.01 & +52:40:41.0 & (Groth strip); DEEP-2, SIRTF, GALEX\\
D4 & 22:15:31.67 & $-$17:44:05.7 & XMM-Deep, GALEX\\
\noalign{\smallskip}
\tableline
\end{tabular}
}
\end{center}
\end{table}

\begin{figure}
\plotone{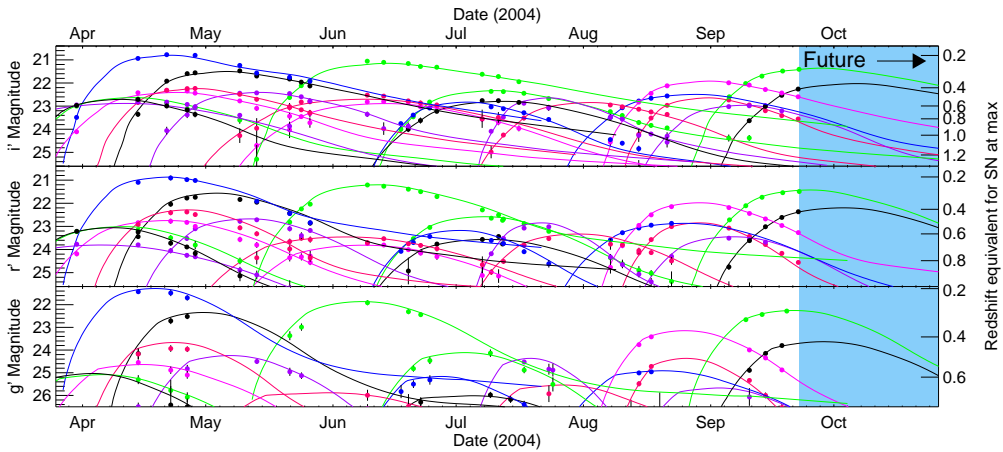}

\caption{
  A selection of real-time light-curves in the $g'r'i'$ filters for
  confirmed SNe\,Ia discovered over the period April--September 2004,
  demonstrating the quality of typical light-curves in the survey. At
  any given lunar phase candidates at maximum light are available for
  follow-up programs. For clarity only around a third of the SNe\,Ia
  followed during this period are shown.\label{rolling-lightcurves}}

\end{figure}

SNLS began in August 2003 (with a pre-survey period from March 2003)
using the queue-scheduled square-degree imager ``megacam'' on CFHT.
In a typical month, each available field (Table~\ref{fieldpositions})
is imaged on five epochs in a combination of $g'r'i'z'$ ($r'i'$ are
always observed; $g'z'$ are arranged according to the lunar phase),
each of the observations spaced 4-5 days apart ($\sim$3 days in the SN
rest-frame). Each field is typically searched for 5 continuous months,
giving around 20 ``field-months'' in a given calendar year -- and
consequently high-quality and continuous light-curves for each SN
candidate lasting many months (Fig.~\ref{rolling-lightcurves}).

On a given night, data is taken, reduced (using the CFHT-developed
data processing
system\footnote{\texttt{http://www.cfht.hawaii.edu/Instruments/Elixir/}}),
aligned, psf-matched and candidates located and placed in our
database\footnote{\texttt{http://legacy.astro.utoronto.ca/}} within 12
hours, allowing a rapid prioritisation for spectroscopic followup. The
spectroscopic time comes via an international collaboration, with
allocations on Gemini (60 hours/semester; see Howell et~al., in prep.
and these proceedings), VLT (60 hours/semester; Basa et~al., in prep.)
and Keck (in the ``A'' semester for coverage of the northern-most D3
Groth Strip field).

The amount of successful spectroscopic follow-up performed will define the 
success of SNLS, and with many hundreds of candidates to select from every 
month, locating probable SNe\,Ia (and rejecting AGN and other variable 
objects) is essential. We have developed a SN photometric redshift 
technique which performs a light-curve fit to 2-3 epochs of multi-band 
real-time data, and returns a probability that the candidate is a SN\,Ia, 
as well as predictions of redshift, stretch and phase. This technique is 
extraordinarily successful; since the technique was implemented our SN\,Ia 
fraction is around 80\%, compared with $\sim$50\% in classical surveys 
(e.g, Lidman et~al. 2004), with an excellent agreement between photometric 
and spectroscopic redshift.

\begin{figure}
  \plottwo{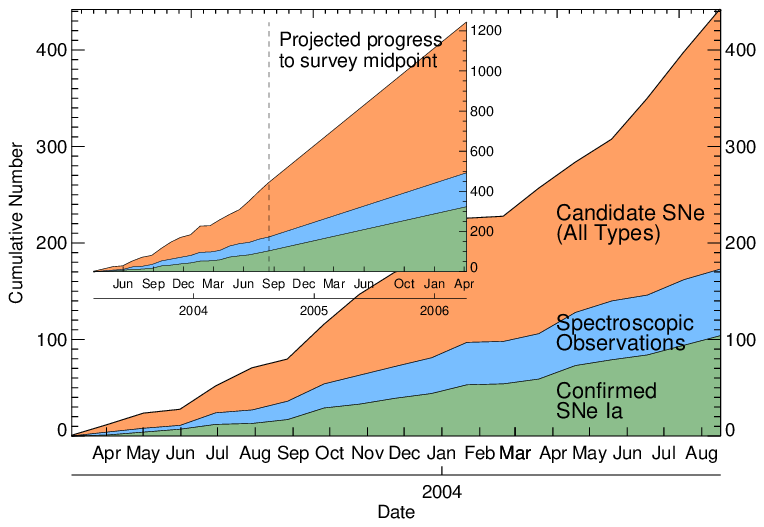}{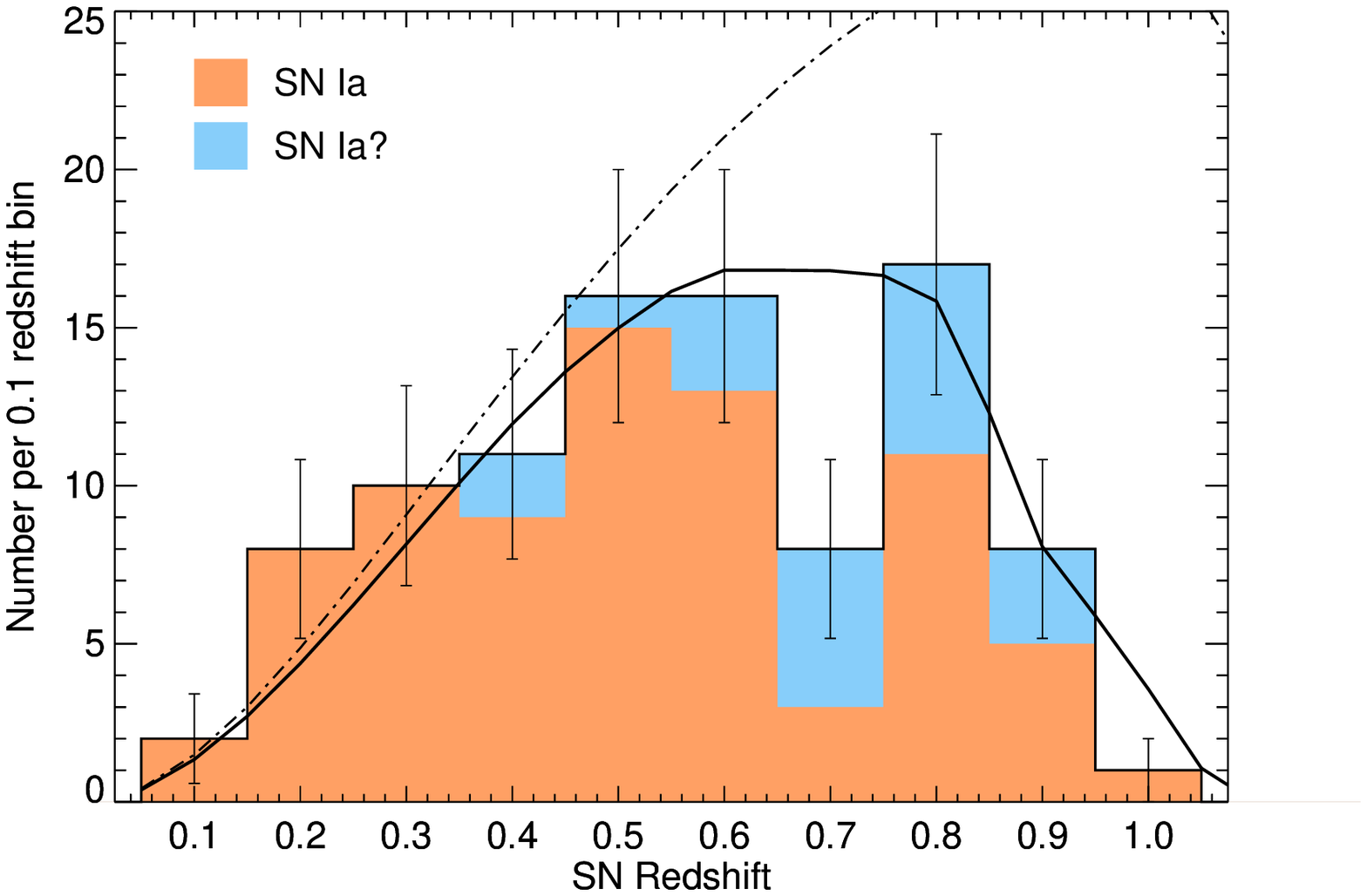}

\caption{LEFT: The survey progress as
  of August 2004. The main panel shows the number of SN candidates,
  spectroscopic observations, and SNe\,Ia for the first year of the
  survey. The inset shows predicted numbers to the survey mid-point
  based on the last three months of the survey. RIGHT: The measured
  $N(z)$ of the SNe\,Ia distribution compared to simple predictions
  based on the search limiting magnitudes and completeness, and the
  spectroscopic efficiencies of the survey.  The dot-dash line shows
  the predicted number of SNe\,Ia candidates detected as a function of
  redshift; the solid line shows the expected distribution of
  confirmed SNe\,Ia after one year based on the amount of
  spectroscopic time available and our follow-up
  strategy.\label{numbercounts}}

\end{figure} 

\section{Current Status}

As of August 2004, SNLS has located $\sim$630 SN candidates, with 110
spectroscopically identified as SNe\,Ia and $\sim$20 core-collapse
SNe. We currently obtain an average of 11 spectroscopically confirmed
SNe\,Ia per month; at this current rate $\sim$700 SNe\,Ia will be
spectroscopically identified over the 5 years of the survey
(Fig.~\ref{numbercounts}, left); this will form the largest and most
homogeneous high-$z$ SN sample available over at least the next
decade. Our current number-redshift ($N(z)$) distribution (together
with some predictions based on published high-redshift SN\,Ia rates
and our spectroscopic efficiency) is shown in Fig.~\ref{numbercounts},
right.

We are currently investigating both the internal photometric alignment
and the external photometric calibration -- both are critical for an
accurate determination of $w$. Internally, after Elixir
data-processing, the 36 chip array is photometrically aligned to
$\sim$0.01\,mag in all 4 filters, and internal colour-terms appear
negligible. As we are continually observing the same four deep fields,
an internal field-to-field calibration of better than 0.01\,mag is
achievable. Beyond the internal calibration, we are in the process of
defining a ``natural megacam'' system onto which we will place our SN
photometry; in the meantime an overlap of the D2 field with the SDSS
data release allows a calibration of our data onto the SDSS system
beyond that obtained by observing SDSS secondary standards.

These 700 well-measured SNe\,Ia, together with an $\Omega_m$ prior
known to $\pm$0.03 (i.e. 10\%), will allow us to determine $w$ to a
statistical precision of $\pm$0.07, distinguishing between
$w$$>$$-0.8$ and $w$$=$$-1$ at 3$\sigma$. Clearly, with 700 SNe\,Ia,
controlling (and understanding) systematics in the SN sample is of the
utmost importance. Our rolling search with multiple filters
($g'r'i'z'$) will generate the first large high-$z$ SN\,Ia dataset
with complete colour coverage throughout the lightcurves (see
Fig.~\ref{rolling-lightcurves}), enabling comprehensive extinction
studies since all the SNe are sampled over a wide, rest-wavelength
baseline. Additionally, the narrow galaxy emission and absorption
lines detectable with spectroscopy of SN+host, plus the extremely deep
ground-based optical coverage due to the rolling-search (and some
\textit{HST} coverage), provide valuable constraints on host galaxy
stellar populations, and allow construction of SN sampled based in
different host galaxy classes (e.g, presumed ``dust-free''
ellipticals).  Thus, SNLS will provide the biggest improvement in the
determination of the dark energy parameters achievable over the next
decade.

\acknowledgements The SNLS collaboration gratefully acknowledges the assistance
of the CFHT Queue Service Observing team.  Canadian collaboration
members acknowledge support from NSERC and CIAR; French collaboration
members from CNRS/IN2P3, CNRS/INSU and CEA.  SNLS relies on
observations with MegaCam, a joint project of CFHT, CEA/DAPNIA and
HIA.

\end{document}